# MARCOT Pathfinder at Calar Alto Progress Report

Martin M. Roth[a], Jesús Aceituno[b,c], José L. Ortiz[c], Kalaga Madhav[a], Stefan Cikota[b], John Davenport[a], Pedro Amado[c,d], Fran Pozuelos[c], Rafael Luque[c], and Nicolas Morales[c]

[a]Leibniz-Institute for Astrophysics Potsdam (AIP), An der Sternwarte 16, 14482 Potsdam, Germany
[b]Centro Astronómico Hispano Aleman, Sierra de los Filabres sn , 04550 Gérgal Almería, Spain
[c]Instituto de Astrofísica de Andalucía, CSIC, Glorieta de la Astronomía SN, 18008, Granada, Spain
[d] Space Sciences, Technologies and Astrophysics Research (STAR) Institute, Université de Liège, 19C Allée du 6 Août, 4000 Liège, Belgium

**ABSTRACT**

MARCOT Pathfinder is a precursor for MARCOT (Multi Array of Combined Telescopes) at Calar Alto Observatory (CAHA) in Spain. MARCOT is intended to provide CARMENES, currently fiber-fed from the CAHA 3.5m Telescope, with a 5-15m light collecting area from a battery of several tens of small telescopes that are incoherently fed into the final joint single fiber feed of the spectrograph. The modular concept, based on commercially available telescopes, results in cost estimates that are a fraction of the ones for extremely large telescopes (ELT). As a novel approach, MARCOT will employ Multi-Mode Photonic Lanterns (MM-PL) that are being developed as a variant of classical photonic lanterns, to combine the light from the individual telescopes to a single fiber feed to the instrument. This progress report presents the overall concept of MARCOT, the pathfinder telescope and enclosure that is being commissioned at CAHA, the concept of MM-PL, and the next step of installing the Potsdam Multiplex Raman Spectrograph (MRS). MARCOT Pathfinder will be used to validate the conceptual design and predicted performance of MM-PL on sky with a 7 unit telescope prototype.

**Keywords:** Very Large Telescope, ground-based telescopes, High resolution spectroscopy

## 1. INTRODUCTION

Ground-based astronomy is currently experiencing the transition from 8-10 m class telescopes to extremely large telescopes (ELT) with primary mirror diameters of up to 39 m [1]. Their science drivers, e.g., cosmology, the study of the formation and evolution of galaxies at high redshift, detailed investigation of resolved stellar populations in nearby galaxies, imagery and spectroscopy of extra-solar planetary systems, etc. demand two major capabilities: (a) high light-collecting power, and (b) diffraction-limited angular resolution, justifying the enormous cost of the new facilities. When in operation, it is expected that there will be a high demand on ELTs, leading to a heavy competition for the sparse observing time available to the community. Considering the over-subscription of currently existing 8-10 m telescopes [2], it is unlikely that ELTs will be made available for highly demanding time domain astronomy. However, three major challenges of these new large facility are (1) manufacture, (2) funding, and (3) the complexity of operation, involving costs on the order of a billion Euro.

However, there is a demand from photon-starved, time-domain science cases that are in need of a large aperture for seeing-limited, low cost observations, requiring typically tens of nights per semester. The MARCOT concept is intended to fill this niche by achieving the light collecting area through the incoherent combination of light from a large number of small telescopes. An example of a similar solution is the Dragonfly Telescope [3], that was developed to image extended structures down to surface brightness levels as low as $\mu B$ = 32 mag arcsec$^{-2}$ in 10 h integrations. No conventional reflecting telescope of whichever size would be capable of reaching this limit. It combines eight commercial telephoto lenses on a common mount that are coupled to eight science grade CCD cameras. Another concept is PolyOculus [4], that proposes to link commercial-off-the-shelf telescopes through a single-mode fiber photonic lantern. Finally, Argus is a concept for a large all-sky telescope, consisting of 900 moderate-aperture (5m-aperture-equivalent), off-the-shelf telescopes multiplexed into a common hemispherical dome to observe each part of the sky for nine minutes at high cadence each night [https://evryscope.astro.unc.edu/the-argus-array/ ].

The MARCOT concept has an effective aperture equivalent to a single 15m telescope, and incoherently combines light from a set of small telescopes or Optical Telescope Assemblies (OTAs). Each OTA is equipped with an acquisition and guiding system and a fiber optic feed, and is built as an individual independent subsystem of the MARCOT telescope. Several OTAs form a module of MARCOT. The light collected by the modules through the individual optical fiber feed of each OTA is optically injected into a novel multi-mode photonic lantern (MM-PL), which can either directly feed the astronomical instrument attached to the telescope or, feed a second stage MM-PL that can combine the light from a cascade of several MARCOT modules. Due to ist modular nature MARCOT can be scaled easily to larger or smaller effective apertures at a fraction of the cost of a classical single primary mirror telescope of the same size.

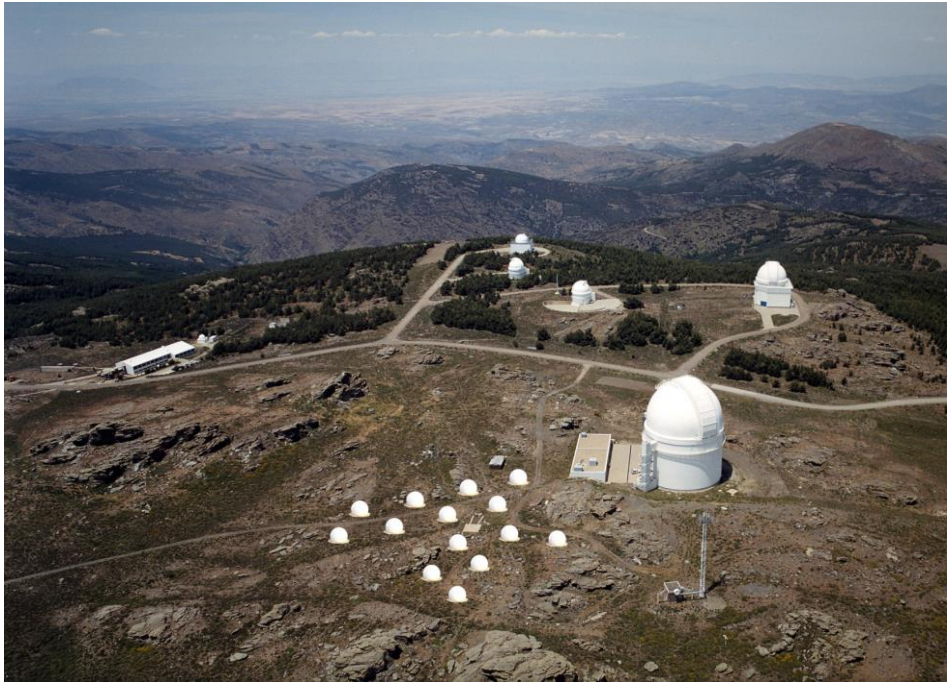

Figure 1. Aerial view of CAHA, with 3.5m dome in front, and a montage of 14 small MARCOT domes to illustrate a possible future configuration. For more information about CAHA, see https://www.caha.es

## 2. MARCOT SCIENCE CASE

The number of frontier scientific research areas to which a large-aperture telescope, with the appropriate instrumentation suite and with enough dedicated observing time, can contribute to is large. However, most are characterized as being photon-starved science cases. Among all, the exoplanet field is particularly interesting, where over the past few decades important discoveries have been made with a number of on-going ground-base (e.g. MEarth [5], HARPS [6] and CARMENES [7, 8], as well as space-based (e.g. Kepler [9] and TESS [10]) surveys. The next generation of missions and large facilities with dedicated instruments to find and characterize exoplanets and planetary systems has the potential to be very fruitful, such as the James Webb Space Telescope (JWST [11]), PLATO [12], the Extremely Large Telescope [13,14], and Ariel [15]. Currently, 5044 exoplanets have been confirmed, a number that is rising on an almost daily basis (https://exoplanets.nasa.gov/). This increasing interest is partially motivated by the opportunity to obtain a complete analysis of exoplanetary statistics to yield important insight into the questions of how typical our Solar System architecture, formation and evolution are. This in turn will help us elucidating the uniqueness of our Solar System in the context of other planetary systems in our Galaxy [16, 17].

The MARCOT concept is chiefly guided by two specific exoplanet science cases:
- Populating the mass-radius diagram
- Exoplanet atmospheres

## 3. MARCOT PATHFINDER TELESCOPE

For the purpose of validating the multiple OTA and MM-PL concept of MARCOT, a single module Pathfinder is currently being built at CAHA. The layout includes 7 OTAs on a common mount, CMOS imagers at each OTA Cassegrain focus, the fiber link with MM-PL, and a broadband multi-channel fiber spectograph.

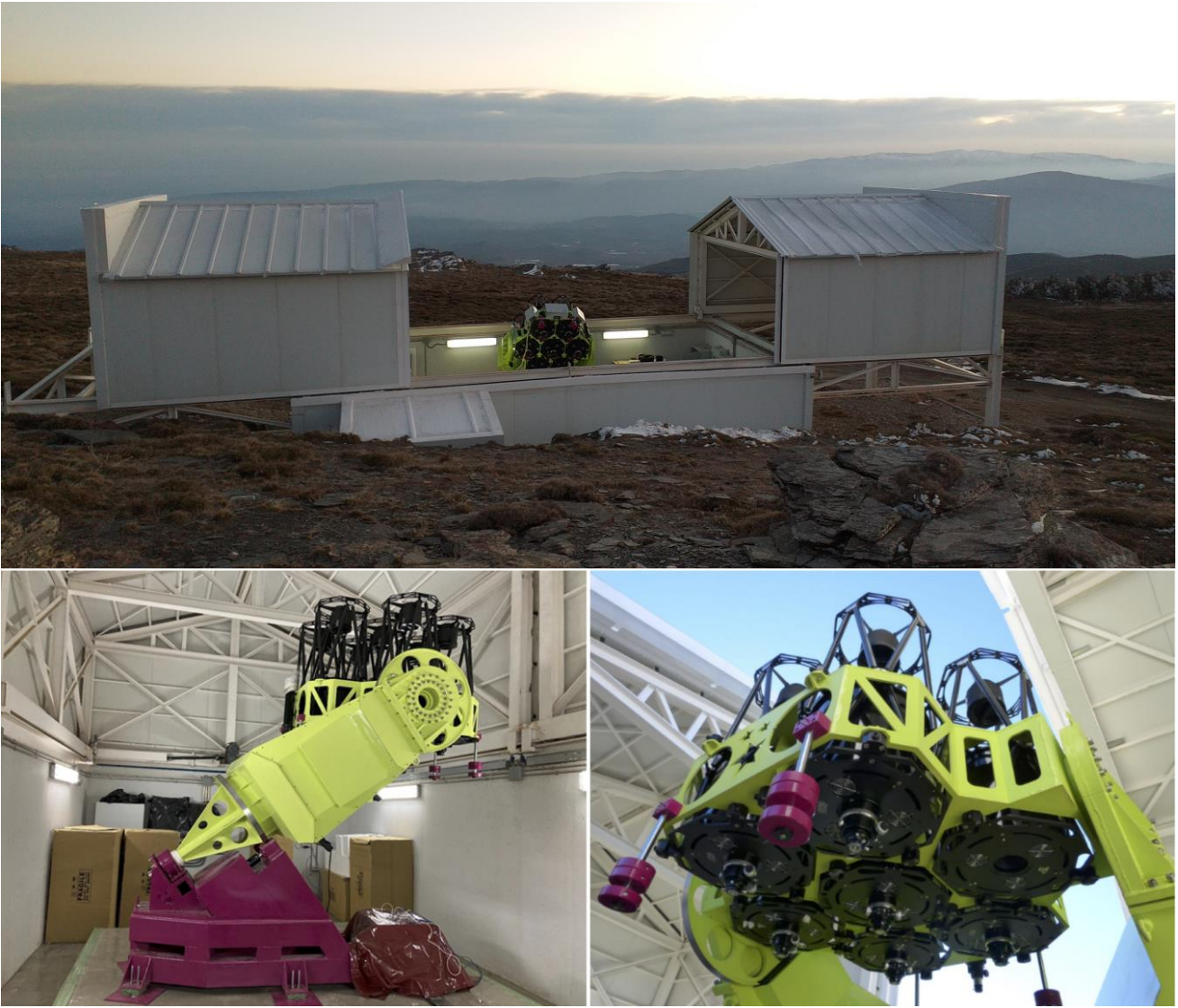

Figure 2. MARCOT Pathfinder with 7 OTAs during commissioning at Calar Alto. Top: retractable roof dome with assembled telescope module. Lower left: side view of the telescope module. Lower right: detail of the main cell.

The optics of the seven off-the-shelf GSO 16" f/8 Ritchey-Chretien Truss Tube OTAs consists of a concave hyperbolic primary and a convex hyperbolic secondary mirror. The mirror is made of quartz and with a highly reflective (99%) dielectric coating. The used imagers are ZWO ASI 1600 GT Mono cameras with internal 5 x 1.25" electronic filter wheels, all equipped with Blue, Green, Red and Luminance photographic filters. The CMOS detectors

(Panasonic MN34230) have a resolution of 4656 x 3520 pixels, with pixel sizes of 3.8 x 3.8 μm$^2$, providing a pixel scale of the OTAs of 0.238 arcsec/pixel, and a field of view of 18.5 x 14 arcmin$^2$. The Pathfinder module is mounted on an equatorial mount, fully designed and manufactured by TecnoHita Instrumentacion (Toledo, Spain).

## 4. MULTI-MODE PHOTONIC LANTERN

The light of a star from each OTA in a module is collected by a single multi-mode fiber (MMF). All the individual MMFs are then tapered into a single MMF, forming a new type of MM-PL, that is then fed to the spectrograph.

At present, the concept is limited to single-object spectroscopy for reasons of complexity. However, multiobject spectroscopy (MOS) or integral field units (IFUs) are not fundamentally ruled out and a topic to be studied in the future. Combining the light from individual fibers for many tens of OTAs into a single fiber feed is the major innovation aspired to with MARCOT.

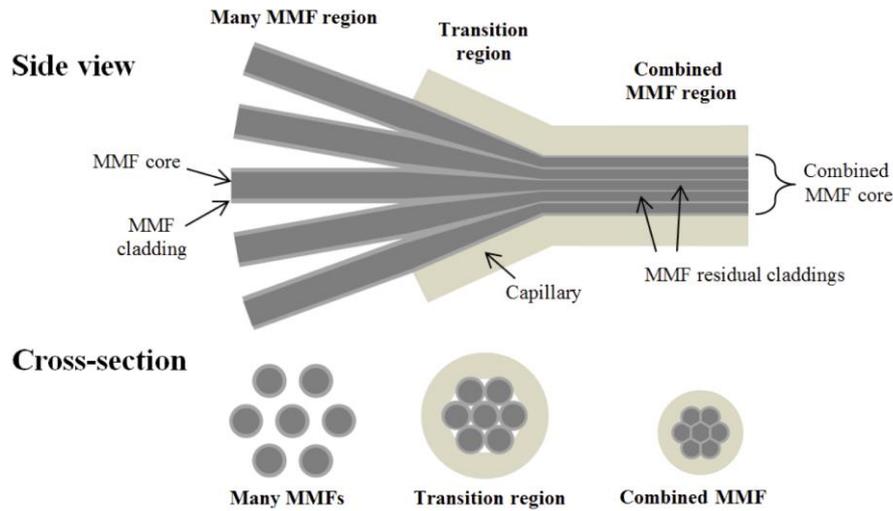

Figure 3. Schematic of an MM-PL made from 7 MMFs. MMFs with high core-to-cladding ratios (left) are tapered down to form a single MM core (right) as the claddings can no longer efficiently confine light. The capillary now forms the cladding of the new combined MMF.

The device that is capable of accomplishing this task is the photonic lantern (PL), which was first put forward by [18]. The classical PL is essentially a mode converter that fuses several input single mode fibers (SMF) into a tapered output MMF. The device also works in the reverse direction such as to distribute light from a single MMF into several SMF. The PL has been shown to work adiabatically with high efficiency. There is an abundant literature for use of PL in telecommunications. In astronomy, the plate scale of large telescopes normally precludes fiber links with SMFs because the core diameter of order 10 μ m would be far too small for the image of a star [19]. Implementing novel photonic functions in astronomical instrumentation therefore requires mode conversion, because such functions usually work only in single mode waveguides. A simple demagnification with a lens does not help since the invariance of etendue of the demagnified beam would overfill the numerical aperture (NA) of the SMF. A good example for mode conversion for photonic functions was presented by [20] for the purpose of OH emission line suppression with Fiber Bragg gratings (FBG), and subsequently validated with the GNOSIS and PRAXIS demonstrators at the Anglo-Australian Telescope [21, 22, 23], showing that the SMF PL delivers an efficiency of 80%. Although the MARCOT OTAs have an aperture of 405 mm and a short focal length of 3250 mm with a plate scale of 63.5 arcsec/mm, which comes close to a typical SMF core diameter for 1 arcsec FWHM seeing, it would be desirable to accommodate a somewhat larger fiber core to collect most of the light from the wings of the stellar point-spread-function, and to reduce the sensitivity to pointing and guiding errors, thus avoiding the cost for high precision acquisition and guiding systems. As the baseline, a fiber core diameter of nominally 0.047 mm is currently envisaged.

This is where the concept of a new type of MMF PL comes into play. Traditionally, a common method for producing PLs is to stack a bundle of single-mode fibers (SMFs) inside a capillary, and taper them down to fuse into a single MMF [24]. The capillary becomes the cladding of the MMF, while the claddings of the SMFs fuse to become the core of the MMF. The cores of the SMFs reduce to the point that they can no longer efficiently contain light. MMFs cannot be drawn into photonic lanterns in the traditional way as their significantly larger cores are still able to couple light even after tapering. This has traditionally meant that PLs could only be used efficiently to combine light from near diffraction limited systems such as extreme AO corrected telescopes, which is unsuitable for MARCOT.

In MM-PL the new core is formed from the original cores instead of the claddings of the individual MMFs. This was done by using reduced-clad MMFs with high core-to-cladding ratios which are then tapered to a point where they are too thin to efficiently confine light. A capillary with a refractive index lower than an individual MMF's cladding is used to ensure a regular arrangement of the MMFs during the tapering process. After tapering, the capillary becomes the new cladding of the resultant combined MMF as shown in Fig. 3 , and the residual claddings of individual MMFs are now too thin to efficiently contain light, effectively functioning as a single large diameter core.

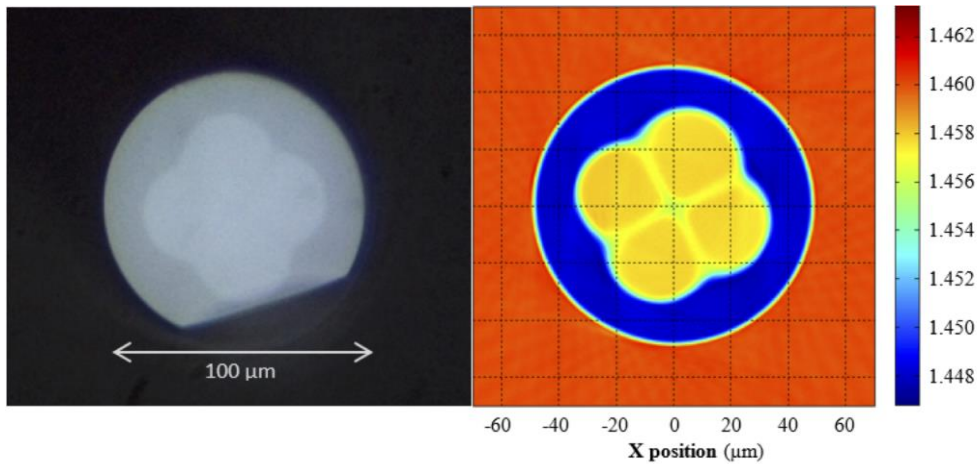

Figure 4. Image of the output end facet of the N=4 MM-PL (left) and its 2D refractive index (RI) profile (right). The four fused cores are visible, but the claddings of the individual fibers are no longer visible in the image but can be seen in the RI image (thin yellow-orange "X" at the centre).

As a first experiment, the fabrication and test of a 4 channel MM-PL was undertaken. The objective was to verify that a simple MM-PL can be manufactured with a performance close to the expectations from numerical simulations, and to establish the relevant test procedures for future development steps. Subsequently, more complex PLs with different core diameters will be manufactured and tested to accommodate the requirements of MARCOT.

The envisaged MM-PL needs to utilize fibers with a thin cladding in order to accomplish an efficient taper that were not readily available off the shelf for the experiment. Instead, an existing CeramOptec OPTRAN-UVWFS with core-to-cladding ration of 1:1.06 was chosen as the base MMF fiber for fabricating a first experimental MM-PLs. This fiber has extremely low loss and can be used over a wavelength range from 200nm to 2000nm, which is ideally suited for MARCOT with CARMENES-VIS and CARMENES-NIR spectrographs. However, the fiber needed to be tapered down to about a tenth of its core diameter in order to reach approximately the envisaged MMF core diameter coming from a first stage PL. For future development of MARCOT Pathfinder MMF-PL, specifically drawn fibers that exhibit already the required diameter and a thin cladding will be used.

To fabricate the PL we selected a fluorine doped capillary with ID=280 μm, OD=390 μm and NA=0.16. A fluorine doped capillary was used because the cores of the base MMF (OPTRAN) were made of silica and therefore a material with lower index was required to confine light from the transition region all the way to the output end of the MM-PL. However, for fabricating N =4 MM-PL, the 1060 μm diameter (cladding) base MMF was too large to bundle 4 such fibers inside the capillary [24] and had to be tapered down. We used a vertical $CO_2$ based glass processing system

(Nyfors SMARTSPLICE) to draw and taper the base MMF down to a suitable diameter. The taper was performed in a multi-step process to minimise irregularities and inhomogeneities, by carefully controlling the $CO_2$ laser power. The base fiber was tapered from 1060 μm → 600 μm → 350 μm → 200 μm and finally to 115 μm, by stepping the $CO_2$ laser power from 7.9% → 5.0% → 4.1% → 3.1% of the maximum power. The diameter of the core after tapering was 108.5 μm.

Four tapered MMF fibers were stacked inside the fluorine capillary after lubricating the interior of the capillary with isopropyl alcohol (IPA). Excess IPA was drawn out using a vacuum pump. The $CO_2$ glass processing system was then used to taper the bundle down from 390 μm → 225 μm → 160 μm → 100 μm, by stepping the $CO_2$ laser power from 4% → 3.3% → 2.9%. The diameter of the core at the output end of N =4 MM-PL was 66 μm and NA=0.16. The microscope image of the cleaved facet of the output end and its refractive index distribution is shown in Fig. 4 . The four fused cores can be seen in the middle surrounded by the fluorine doped capillary that forms the cladding and the original claddings of the individual MMFs are not visible. The multiple step tapering process has reduced the cladding thickness by a factor of 36 down to 0.83 μm, so it is expected that they can no longer be seen or, confine light.

## 5. MARCOT PATHFINDER SPECTROGRAPH

Instead of interrupting ongoing CARMENES surveys for extensive tests and optimization of the MARCOT fiber link, the Pathfinder will be equipped with a dedicated spectrograph in the same dome. We will use the existing fiber-fed Multiplex-Raman-Spectrograph (MRS) that was initially built for a now completed technology transfer project at AIP [25,26]. Key parameters of this instrument are summarized in Table 1.

The Pathfinder will allow unlimited on-sky tests of all critical components, including a detailed end-to-end performance study using the mode expansion theory technique of [27] with fibers of different core diameters and arbitrary lengths. The MRS spectrograph has an experimental slot to provide access to the diverging beam that emerges from the first collimator lens which allows a direct measurement of the far field of the attached fibers for in-situ FRD measurements, including agitation. The capacity of > 400 fibers on the pseudo-slit is more than enough to couple one PL output and 7 conventional MMF in parallel, which enables a dual fiber pickup at each OTA, i.e. the possibility of a direct differential test between PL (A) and conventional MMF (B) by off-setting the telescope and A-B-A throughput measurements. The facility will be used extensively to develop and optimize the MARCOT fiber link, especially testing different generations of MM-PL and different configurations/lengths of few mode fibers, for final implementation at CARMENES. Also alternative solutions, such as microlens coupling, can be investigated.

| Optical System | fully refractive collimator and camera |
| --- | --- |
| Grating | Volume Phase Holographic Grating |
| Detector | e2v CCD 231, 4Kx4K 15 µm pixels |
| Collimator focal ratio | f/4.33 |
| Useable pseudo-slit length | 118 mm |
| Spectral resolution (with 50 um fiber) | R = 1200 . . . 3000 (blue to red) |
| Wavelength range | 450 .. 900 nm |
| Ensquared energy in 2x2 pixels (measured) | > 82% over all field angles and wavelengths, typically 90% |

Table 1. Parameters of MRS Spectrograph

## 6. MARCOT PATHFINDER STATUS

The MARCOT Pathfinder telescope has been delivered and installed at the existing retractable dome at CAHA in early 2022. Fig. 5 shows a photograph of the First Light configuration, where each OTA Cassegrain focus was equipped with an individual CMOS camera in order to assess the alignment and stability of the system. First Light was obtained on January 24, 2022. An immediate discovery was a shortcoming of the off-the-shelf OTAs in that all of the primary mirror supports were found to be highly unstable, due to an inadequate fixture, i.e. a feature of the commercial telescopes. A solution to this problem is straight forward and is being solved.

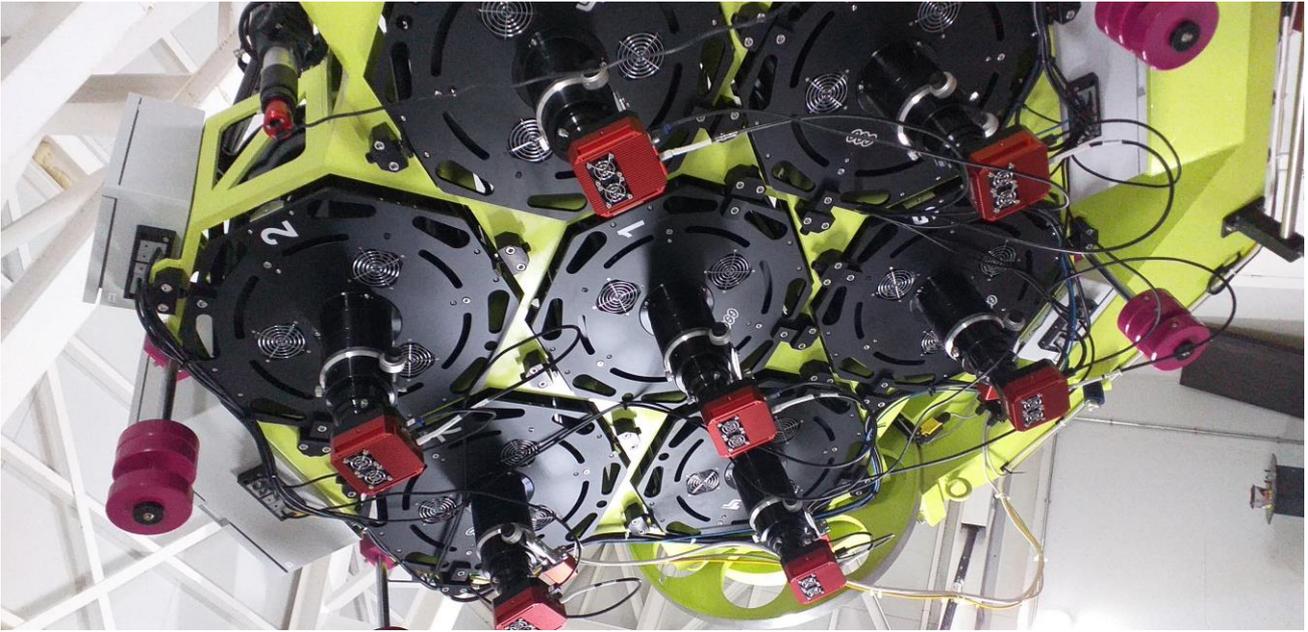

Figure 5. MARCOT Pathfinder main cell with 7 OTAs, each equipped with a thermoelectrically cooled CMOS camera.

Despite the OTA instability issue, short observation series were obtained to test the capability of reprojecting and combining of FITS images collected with the 7 OTAs simultaneously.

The reprojection of FITS images was performed by using Python's "reproject" package [28]. The reproject package implements image reprojection (resampling) methods for astronomical images and more generally ndimensional data. These assume that the WCS information contained in the FITS headers is correct. The package consists of a few high-level functions to do reprojection using different algorithms. In the analysis example shown below, the function "reproject interp()" was used, which reprojects data to a new projection using interpolation, and which is typically the fastest way to reproject an image.

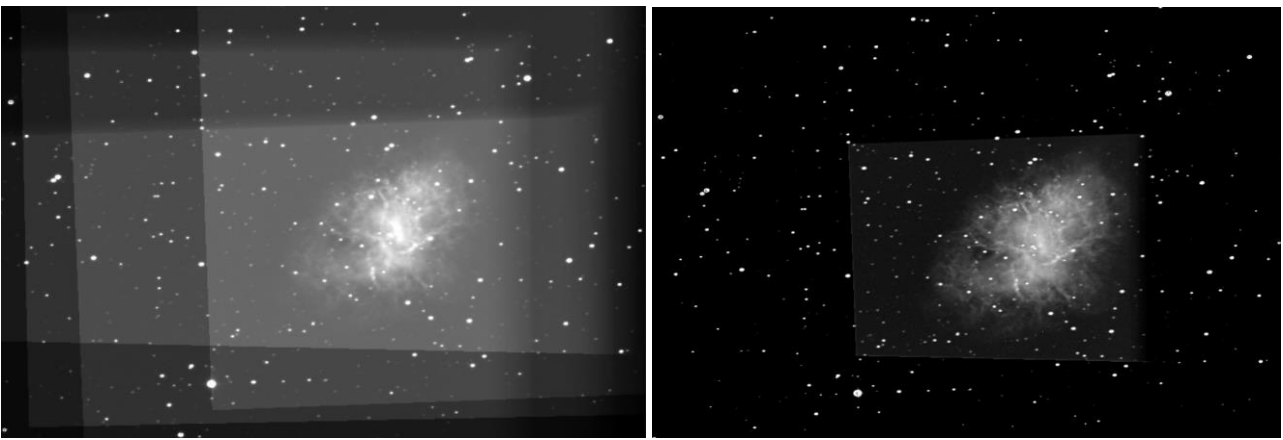

Figure 6. Right: Composite image of Messier 1 taken on January 29, 2022, at 20:45 UT and containing 399 images in total. With each of the Pathfinder's seven OTAs, 57 images with an integration time of 60 seconds were acquired, reprojected and coadded. Left: Same composite as shown right, but with a different set of background and dynamic range, so that the overlap of the individual reprojected images is better discernible.

Fig. 6 on the right-hand side shows a composite of coadded images of Messier 1 taken on January 29, 2022, at 20:45 UT and containing 399 images in total. With each of the Pathfinder's seven OTAs, 57 images with 23940 seconds (399 minutes). At the time of the observations, Pathfinder's seven optical tubes were still not fully aligned. Therefore the overlap of the individual images is only ∼ 30%. Fig. 6 on the left-hand side shows the same composite as shown in the right-hand side, but with a different set of background and dynamic range, so that the overlap of the individual reprojected images is better discernible. Fig. 7 illustrates a similar observation of NGC3628. an integration time of 60 seconds were acquired. The total integration time of all images in the composite is 23940 seconds (399 minutes). At the time of the observations, Pathfinder's seven optical tubes were still not fully aligned. Therefore the overlap of the individual images is only ∼ 30%. Fig. 6 on the left-hand side shows the same composite as shown in the right-hand side, but with a different set of background and dynamic range, so that the overlap of the individual reprojected images is better discernible. Fig. 7 illustrates a similar observation of NGC3628.

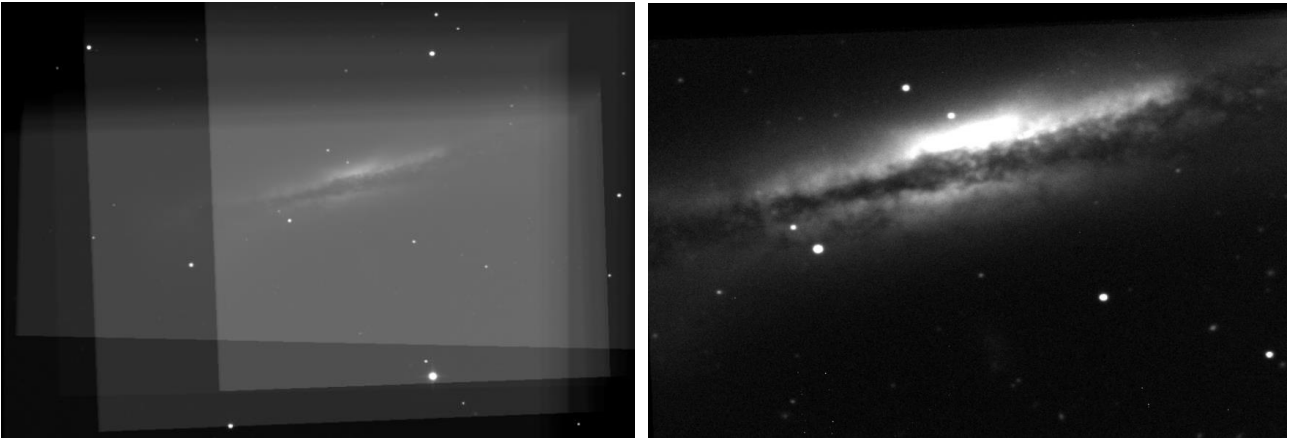

Figure 7. Composite image of NGC 3628, containing 210 images in total, taken with MARCOT Pathfinder on January 29, 2022 in the same configuration as described above. 30 images with an integration time of 60 seconds were acquired. The total integration time of all images in the composition is 12600 seconds (210 minutes). Again, the overlap of the individual images was only 30%. Left: raw composition of contributing images. Right: reprojected final image.

## 7. CONCLUSIONS AND NEXT STEPS

The installation of MARCOT Pathfinder is an important milestone in the process of validating the MARCOT concept on the roadmap towards a real implementation. The commissioning process is ongoing, awaiting the arrival of the MRS spectrograph, and the installation of a first classical fiber link, consisting of seven individual fibers. This configuration will be used to establish a solid understanding of the fiber feed performance, reveal any issues of focusing, guiding, etc., and also provide a platform for data reduction and analysis. In parallel, the development of a first prototype of a 7 channel MM-PL will be undertaken to perform the first critical on-sky tests of this device. After review of the test results from the prototype, the development of next generation MM-PL toward an optimized configuration is foreseen to enable the launch of the full-fledged MARCOT project.

## ACKNOWLEDGEMENTS

The authors acknowledge financial support from the State Agency for Research of the Spanish MCIU through the Center of Excellence Severo Ochoa award to the Instituto de Astrofısica de Andalucıa (SEV20170709). This research has been partially funded by the Junta de Andalucıa (SOMM17 5208 IAA).

KM, JD, and MMR acknowledge support from BMBF grant 03Z22AN11 "Astrophotonics", DFG grant 326946494, "NAIR", and BMBF grant 03Z22AI1 "Strategic Investment", at the Zentrum fürr Innovationskompetenz innoFSPEC.

# REFERENCES

[1] Gilmozzi, R., "Science and technology drivers for future giant telescopes" in Ground-based Telescopes, Oschmann, Jacobus M., J., ed., Society of Photo-Optical Instrumentation Engineers (SPIE) Conference Series **5489**, 1–10 (Oct. 2004)

[2] Grothkopf, U. and Meakins, S., Basic ESO Publication Statistics, European Southern Observatory (ESO), Garching bei München, Germany (2021)

[3] Abraham, R. G. and van Dokkum, P. G., "Ultra-Low Surface Brightness Imaging with the Dragonfly Telephoto Array", PASP **126**, 55 (Jan. 2014).

[4] Moraitis, C. D., Alvarado-Zacarias, J. C., Amezcua-Correa, R., Jeram, S., and Eikenberry, S. S., "Demonstration of high-efficiency photonic lantern couplers for polyoculus", Appl. Opt. **60**, D93–D99 (Jul 2021).

[5] Berta, Z. K., Irwin, J., Charbonneau, D., Burke, C. J., and Falco, E. E., "Transit Detection in the MEarth Survey of Nearby M Dwarfs: Bridging the Clean-first, Search-later Divide", AJ **144**, 145 (Nov. 2012).

[6] Mayor, M. et al., "Setting New Standards with HARPS", The Messenger **114**, 20–24 (Dec. 2003).

[7] Quirrenbach, A. et al. "CARMENES instrument overview" in Society of Photo-Optical Instrumentation Engineers (SPIE) Conference Series **9147**, 91471F (Jul 2014).

[8] Quirrenbach, A. et al., "CARMENES: high-resolution spectra and precise radial velocities in the red and infrared" in Society of Photo-Optical Instrumentation Engineers (SPIE) Conference Series **10702**, 107020W (Jul 2018).

[9] Borucki, W. J. et al., "The Kepler mission: a wide-field-of-view photometer designed to determine the frequency of Earth-size planets around solar-like stars" in Future EUV/UV and Visible Space Astrophysics Missions and Instrumentation., Blades, J. C. and Siegmund, O. H. W., eds., Society of Photo-Optical Instrumentation Engineers (SPIE) Conference Series **4854**, 129–140 (Feb. 2003).

[10] Ricker, G. R. e. a., "Transiting Exoplanet Survey Satellite (TESS)", Journal of Astronomical Telescopes, Instruments, and Systems **1**, 014003 (Jan. 2015).

[11] Gardner, J. P. e. a., "The James Webb Space Telescope", SSR **123**, 485–606 (Apr. 2006).

[12] Rauer, H. e. a., "The PLATO 2.0 mission", Experimental Astronomy **38**, 249–330 (Nov. 2014).

[13] Gilmozzi, R. and Spyromilio, J. "The European Extremely Large Telescope (E-ELT)", The Messenger **127**, 11 (Mar. 2007).

[14] Rodler, F., Exoplanet Research in the Era of the Extremely Large Telescope (ELT), Handbook of Exoplanets **194**, Springer, Cham (2018).

[15] Tinetti, G. e. a., "A chemical survey of exoplanets with ARIEL", Experimental Astronomy **46**, 135–209 (Nov. 2018).

[16] Limbach, M. A. and Turner, E. L., "Exoplanet orbital eccentricity: Multiplicity relation and the Solar System", Proceedings of the National Academy of Science **112**, 20–24 (Jan. 2015).

[17] Horner, J., Kane, S. R., Marshall, J. P., Dalba, P. A., Holt, T. R., Wood, J., Maynard-Casely, H. E., Wittenmyer, R., Lykawka, P. S., Hill, M., Salmeron, R., Bailey, J., Löhne, T., Agnew, M., Carter, B. D., and Tylor, C. C. E., "Solar System Physics for Exoplanet Research", PASP **132**, 102001 (Oct. 2020).

[18] Leon-Saval, S. G., Birks, T. A., Bland-Hawthorn, J., and Englund, M., "Multimode fiber devices with single-mode performance", Optics Letters **30**, 2545–2547 (Oct. 2005).

[19] Diab, M., Dinkelaker, A. N., Davenport, J., Madhav, K., and Roth, M. M., "Starlight coupling through atmospheric turbulence into few-mode fibres and photonic lanterns in the presence of partial adaptive optics correction", MNRAS **501**, 1557–1567 (Feb. 2021).

[20] Bland-Hawthorn, J., Ellis, S. C., Leon-Saval, S. G., Haynes, R., Roth, M. M., Löhmannsröben, H. G., Horton, A. J., Cuby, J. G., Birks, T. A., Lawrence, J. S., Gillingham, P., Ryder, S. D., and Trinh, C., "A complex multi-notch astronomical filter to suppress the bright infrared sky", Nature Communications **2**, 581 (Dec. 2011).

[21] Trinh, C. Q., Ellis, S. C., Bland-Hawthorn, J., Horton, A. J., Lawrence, J. S., and Leon-Saval, S. G., "The nature of the near-infrared interline sky background using fibre Bragg grating OH suppression", MNRAS **432**, 3262–3277 (July 2013).

[22] Ellis, S. C., Bland-Hawthorn, J., Lawrence, J., Horton, A. J., Trinh, C., Leon-Saval, S. G., Shortridge, K., Bryant, J., Case, S., Colless, M., Couch, W., Freeman, K., Gers, L., Glazebrook, K., Haynes, R., Lee, S., Löhmannsröben, H. G., O'Byrne, J., Miziarski, S., Roth, M., Schmidt, B., Tinney, C. G., and Zheng, J.,

[22] "Suppression of the near-infrared OH night-sky lines with fibre Bragg gratings - first results", 425, 1682–1695 (Sept. 2012).

[23] Ellis, S. C., Bland-Hawthorn, J., Lawrence, J. S., Horton, A. J., Content, R., Roth, M. M., Pai, N., Zhelem, R., Case, S., Hernandez, E., Leon-Saval, S. G., Haynes, R., Min, S. S., Giannone, D., Madhav, K., Rahman, A., Betters, C., Haynes, D., Couch, W., Kewley, L. J., McDermid, R., Spitler, L., Sharp, R. G., and Veilleux, S., "First demonstration of OH suppression in a high-efficiency near-infrared spectrograph", MNRAS **492**, 2796–2806 (Feb. 2020).

[24] Davenport, J. J., Diab, M., Madhav, K., and Roth, M. M., "Optimal SMF packing in photonic lanterns: comparing theoretical topology to practical packing arrangements", Journal of the Optical Society of America B Optical Physics **38**, A7 (July 2021).

[25] Moralejo, B., Roth, M. M., Godefroy, P., Fechner, T., Bauer, S. M., Schmälzlin, E., Kelz, A., and Haynes, R., "The Potsdam MRS spectrograph: heritage of MUSE and the impact of cross-innovation in the process of technology transfer" in Advances in Optical and Mechanical Technologies for Telescopes and Instrumentation II, Navarro, R. and Burge, J. H., eds., Society of Photo-Optical Instrumentation Engineers (SPIE) Conference Series **9912**, 991222 (July 2016).

[26] Schmälzlin, E., Moralejo, B., Gersonde, I., Schleusener, J., Darvin, M. E., Thiede, G., and Roth, M. M., "Nonscanning large-area Raman imaging for ex vivo/in vivo skin cancer discrimination", Journal of Biomedical Optics **23**, 105001 (Oct. 2018).

[27] Hernandez, E., Roth, M. M., Petermann, K., Kelz, A., Moralejo, B., and Madhav, K., "Mode expansion theory and application in step-index multimode fibers for astronomical spectroscopy", Journal of the Optical Society of America B Optical Physics **38**, A36 (July 2021).

[28] Robitaille, T., Deil, C., and Ginsburg, A., "reproject: Python-based astronomical image reprojection." Astrophysics Source Code Library, record ascl:2011.023 (Nov. 2020).